\documentclass{elsart}

\usepackage{natbib}
\usepackage{graphicx}
\usepackage{amssymb}

\newcommand{\n}{\'{n}}
\newcommand{\lm}{\lambda}

\begin{document}

\begin{frontmatter}

\title{Impact of the Atmospheric Refraction on the Precise Astrometry\\
           with Adaptive Optics in Infrared}

\author{Krzysztof G. He{\l}miniak}

\address{Nicolaus Copernicus Astronomical Center (NCAC), Polish Academy of Sciences, Rabianska 8, 87-100 Toru\n, Poland\ead{xysiek@ncac.torun.pl}}

\begin{abstract}
We study the impact of the atmospheric differential chromatic refraction
on the measurements and precision of relative astrometry. Specifically, we 
address the problem of measuring the separations of close pairs of binary 
stars with adaptive optics in the J and K bands.

We investigate the influence of 
weather conditions, zenithal distance, star's spectral type and observing 
wavelength on the astrometric precision and determine the accuracy of these 
parameters that is necessary to detect exoplanets with existing and planned 
large ground based telescopes with adaptive optics facilities.
The analytical formulae for simple monochromatic refraction and a full 
approach, as well as moderately simplified procedure, are used to compute 
refraction corrections under a variety of observing conditions.

It is shown that the atmospheric refraction must be taken into account in 
astrometric studies but the full procedure is not necessary in many cases. 
Requirements for achieving a certain astrometric precision are specified.

\end{abstract}


\end{frontmatter}

\section{Introduction}

Recently, the atmospheric refraction (AR) has been subject of several studies
about its impact on the observations \citep{roe02}, the determination of AR at
various wavelength \citep[eg.][]{kun05} or its theoretical models
\citep[eg.][]{gar67, yat95}. The simplest yet quite precise model of the 
atmosphere and the refraction assumes spherical symmetry of the Earth and the 
dependence on local weather conditions \citep[eg.][]{gre85}. The refraction 
decreases the real zenithal distance of an object $z_t$. The refraction angle 
$R=z_t - z_a$, where $z_a$ denotes the apparent observed zenithal distance, is 
highly dependent on many factors such as the observing wavelength, 
air pressure and temperature, humidity and $z_t$ itself. In infrared it usually 
reaches tenths of arcseconds. For relative astrometry, it means that AR 
changes the apparent separation of two stars at two different zenithal 
distances by $R_{21} = |R_2 - R_1|$ along the direction to the zenith. 
The smaller the difference between the zenithal distances of stars, the 
smaller $R_{21}$. Nevertheless, even for separations of several arcseconds, 
AR's contribution to the apparent separation can be larger than the precision 
of an astrometric measurement. Clearly, the impact of AR on the relative 
astrometry of close pairs deserves to be studied carefully.

Modern adaptive optics (AO) systems allow us to obtain diffraction-limited 
images of stars. Sharp and well sampled images are the key to achieving 
$\mu as$-precision which in case of close binaries means an ability to detect 
massive planets around their components. A precision below 
100 $\mu as$ was already achieved with the 8-m VLT \citep{neu06}
and 200-in Hale telescope \citet{cam08}. Future large and 
extremely large telescopes like the Thirty-Meter Telescope, Giant Magellan 
Telescope or the European Extremely Large Telescope, 
equipped with a new-generation extreme AO (ExAO) should reach a 
level of astrometric precision of 10 $\mu as$ or better. Such a precision 
is sufficient to astrometrically detect the movement of a 1 $M_{\odot}$ star 
10 $pc$ away around a common mass-center with a body of $\sim 0.16$ Jupiter 
masses on a 1 $AU$ orbit\footnote{See Equation 4.1 in \citet{hel08}}. 
As we will demonstrate, the relative astrometry at this level of precision will
require very accurate knowledge of meteorological conditions near the telescope.

\section{Modeling AR}

\subsection{Refractive index $n$}
It is not straightforward to derive an analytic formula for a refractive
index $n$. A relatively simple but useful approximation is given by
\citet{roe02} who uses and corrects \citet{sch00}:
\begin{eqnarray}\label{rel_nroe}
\begin{array}{ll}
n(\lm,p,T,p_w) & = 1 \\ 
& + \left[64.328+\frac{29498.1}{146-\lm^{-2}}+\frac{255.4}{41-\lm^{-2}}\right]
\frac{pT_s}{p_sT}10^{-6}\\
 & -43.49\left[1 - \frac{0.007956}{\lm^2}\right]\frac{p_w}{p_s}10^{-6},
\end{array}
\end{eqnarray}
where the observing wavelength, $\lm$, is given in $\mu m$, \textit{p, T} and 
$p_w$ are the pressure $[hPa]$, temperature $[K]$ and partial pressure of 
water vapor $[hPa]$ respectively. The symbols with the index $s$ refer to the 
standard values of air pressure (1013.25 $hPa$) and temperature (288.15 $K$). 

\citet{sch00} computed values of $n$ for the range of wavelengths from 0.2 to 10
$\mu m$. This regime contains many regions, where the presence of atmospheric 
CO$_2$ and water vapor lines cause fluctuations in the refractive index \citep{mat04}.
These so called resonances make the dependence of $n(\lm)$ not so simple as 
\citet{sch00} claim. It is especially important
for the K band which "redder" side is strongly influenced by a resonance with
water at $\sim 2.6\, \mu m$.

This particular resonance is also not included in the model proposed by 
\citet[][with further supplements]{cid96} which now is considered as the
state-of-the-art and is recommended for geological and astronomical research.
This model is based on a revised equation for the density of moist air (with CO$_2$),
known as the BIPM 1981/91 equation \citep{dav92} and assumes that the atmosphere
is a mixture of "dry air", containing a variable amount of carbon-dioxide and 
water vapor. The entire recipe for calculating the refractive index\footnote{In
\citet{cid96} named the "phase" index, not the "group" one} is rather 
complicated \citep[Appendix B in:][]{cid96}. The validity of Ciddor's 
model extends from 0.3 to 1.7 $\mu m$ and from 100 to 1400 $hPa$. 
It means that we need to extrapolate it to the K band ($\sim 2.2\, \mu m$) without 
any warranty of validity, but it covers lower air pressures typical for 
high-altitude observatories.

The last model we considered was presented by \citet{mat04}. It is based on 
the calculations of a complex-valued dielectric function $\varepsilon$ 
(where $n = \Re{(\sqrt{\varepsilon})}$) as the response of a superposition of
independent molecular oscillators whose strengths were derived from the HITRAN
database \citep{rot98}. Almost 60,000 H$_2$O and CO$_2$ lines between 0.44 and
25 $\mu m$ were incorporated to ensure that the influence of the resonances. 
For the results obtained outside the resonances, the following smooth polynomial 
was fitted \citep{mat07}:
\begin{equation}\label{rel_mat}
n - 1 = \sum_{i=0,5} c_i(T,p,H) (\nu - \nu_{ref})^i
\end{equation}
\begin{eqnarray}\label{rel_mat2}
\begin{array}{lcl}
c_i(T,p,H) & = & c_{iref}\\
& + & c_{iT} ( 1/T - 1/T_{ref} ) + c_{iTT} ( 1/T - 1/T_{ref} )^2\\
& + & c_{iH} ( H - H_{ref} ) + c_{iHH} ( H - H_{ref} )^2\\
& + & c_{ip} ( p - p_{ref} ) + c_{ipp} ( p - p_{ref} )^2\\
& + & c_{iTH} ( 1/T - 1/T_{ref} )( H - H_{ref} )\\
& + & c_{iTp} ( 1/T - 1/T_{ref} )( p - p_{ref} )\\
& + & c_{iHp} ( H - H_{ref} )( p - p_{ref} ),
\end{array}
\end{eqnarray}
where $H$ denotes the relative humidity (in \%), $\nu = 1/ \lm$ is the wavenumber
and reference values for $T$, $p$ and $H$ are set to 290.65 $K$, 75,000 $Pa$ and
10 \% respectively. Values of all $c$ coefficients and $\nu_{ref}$ are dependent 
on the wavelength range and for 1.3 to 2.5 $\mu m$ are given in Table I in
\citet{mat07}. This range also limits the validity of the fit.

\begin{figure}
   \centering
   \includegraphics[width=8cm, angle=270]{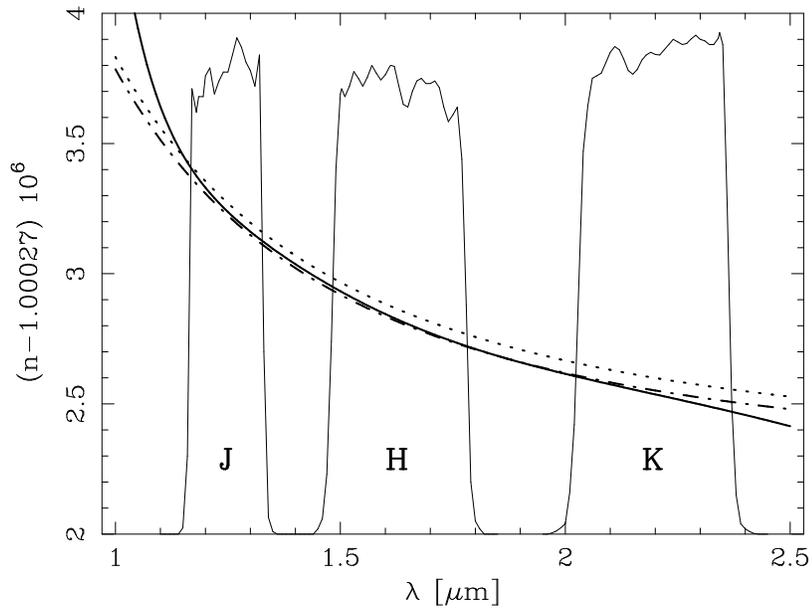}
      \caption{The refractive index as a function of wavelength for the 
standard temperature, pressure and 50\% relative humidity, overplotted 
with the filter transmission curves of the PHARO camera (thin solid line). 
Roe's, Ciddor's and Mathar's models are plotted with thick solid, dashed 
and dot-dashed lines respectively. 100\% transmission is at 
$(n-1.00027) = 4 \times 10^{-6}$.\label{fig_n}}
   \end{figure}

In order to compare the three models above, we plot in Figure \ref{fig_n} the 
refractive index as a function of wavelength. For all the cases the conditions 
are $p=1013.25\, hPa$, $T = 288.15\, K$, 50\% of relative humidity and 375 
ppm (particles per million) of CO$_2$ (not present in Roe's model).
The transmission curves of J, H and K filters of the {\it Palomar High Angular
Resolution Camera} \citep[PHARO][]{hay01} are overplotted. 

The Ciddor's model, considered here as the reference one, produces values 
significantly higher by about $5 \times 10^{-8}$ 
than Roe's over the whole range. It may be due to the fact that Roe's model does not 
include CO$_2$. Nevertheless, the Mathar's model is in excellent agreement with Roe's 
up to $\sim 2.1\, \mu m$ where the previously mentioned resonance with water plays 
a significant role. The majority of the J band is out of Mathar's model validity
range so the curve was extrapolated. As it was shown in \citet{mat07}, this model 
exceeds the measurements of the refractive index of moist air by an almost 
constant value of $4 \times 10^{-8}$ except for the resonance region where 
the empirical data is smaller by $6 \times 10^{-8}$.  

For the remaining calculations, we decided to reject the Ciddor's model as the harder 
to do, and use the two others: Roe's model as the simple one and Mathar's as the 
one which includes resonances, thus the more accurate one. Their accuracy is well 
enough for this application.

Let us note that in order to make the influence of the refraction more predictable 
in general, it is better to observe in the infrared. For $\lm \sim 0.5\mu m$, the 
refraction index is a much steeper function of wavelength than even for J band 
($1.25 \mu m$) while in the K band ($2.2\mu m$) can be considered as almost constant 
\citep[see Fig. 2 in:][]{roe02}. Obviously, also for the adaptive optics purposes it 
is better to operate in the longer wavelengths. We have also found it interesting to 
explore the accuracy of a simplified model in the presence of the water 
vapor resonance. Thus only the K band was chosen for further calculations.

\subsection{Refraction angle and relative astrometry}
Deriving the relation between the refractive index $n(\lm, p, T, p_w)$ 
(or $H$ instead of $p_w$) and the refraction angle $R(n, z_t)$ is not 
straightforward either. The relation for the refraction angle of a monochromatic 
beam of light, $R_{mon}$, proposed by \citet{sch00} and \citet{roe02}
is:
\begin{equation}\label{rel_Rmon}
R_{mon}\,[as] \equiv z_t - z_a \simeq 206265\left(\frac{n^2 -1}{2n^2}\right) \tan z_t ,
\end{equation}
A more sophisticated approach is presented by \citet{sto96} where $R_{mon}$
in the visible (VIS) depends on $\tan^3z_t$ and the non-spherical shape of the 
Earth is taken into account. \citet{sto96} also presents a simple way to 
compute the mean refraction $R_m$ by weighting the individual refractions 
$R_{mon}(\lm)$ with the apparent stellar flux at the wavelength $\lm$ and by
averaging across the bandpass:
\begin{equation}\label{rel_sto}
R_{m}=\frac{\int^{\infty}_0 S(\lm)E(\lm)A(\lm)L(\lm)F(\lm)D(\lm)R_{mon}(\lm)d\lm}
{\int^{\infty}_0 S(\lm)E(\lm)A(\lm)L(\lm)F(\lm)D(\lm)d\lm},
\end{equation}
where $S(\lm)$ is the spectral energy distribution of a star, $E(\lm)$ -- 
the transmittance of interstellar dust, $A(\lm)$ -- transmission of the 
atmosphere at a given airmass, $L(\lm)$ -- the transmission of the telescope 
optics, $F(\lm)$ -- the filter transmission and $D(\lm)$ -- the quantum 
efficiency of the detector. This averaging again favors the IR, where 
the method by Stone approaches the one by Roe. 

\begin{figure*}
   \centering
   \includegraphics[width=0.5\textwidth]{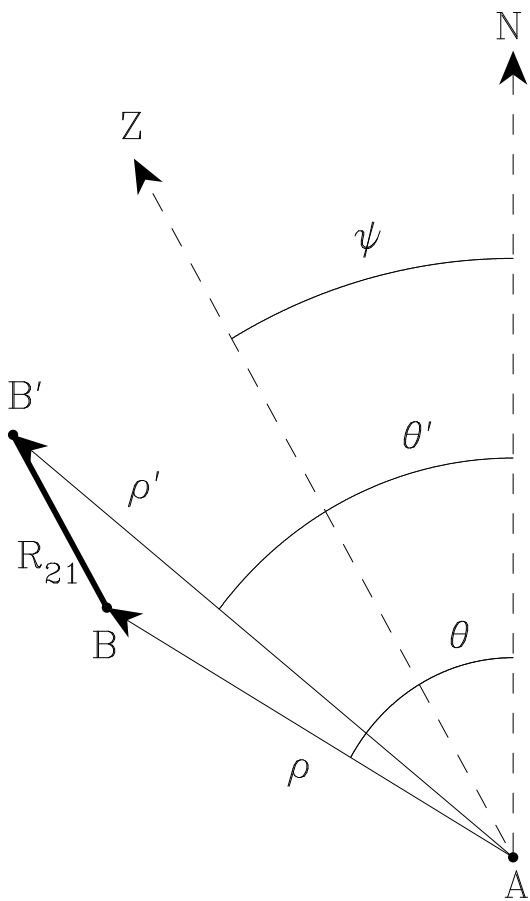}
      \caption{Geometry of the refraction and its impact on the relative 
astrometry of a close pair of stars. N is the direction to the North, Z to 
the Zenith. The second star is observed at the point B ($\rho$, $\theta$), 
while its real location is at B'($\rho'$, $\theta'$), {\it relatively} to A.
\label{fig_arc}}
   \end{figure*}

Figure \ref{fig_arc} depicts the influence of AR on relative astrometry of 
binaries or other close pairs. AR changes the separation between the objects 
by a value of $R_{21} = |R_{m1} - R_{m2}|$, along the direction to the Zenith. 
From the geometry of the effect, the following relation can be derived:
\begin{equation}\label{rel4}
\rho'^2 = \rho^2 + R_{21}^2 + 2\rho R_{21}\cos(\theta - \psi),
\end{equation}
where $\rho'$, $\rho$ are respectively the true and the apparent separations,
$\theta'$ and $\theta$ are respectively the true and the apparent 
position angles of the second star, both measured from the vector 
pointing to the North counter-clockwise to the position vector of the star 
B relatively to A, and $\psi$ is the paralactic angle (between the 
North and Zenith). 

One has also keep in mind that during a single observation, the zenithal distance
of the system changes with time, thus, in general, also the relative observed
separation. Most of the Adaptive Optics systems guide in VIS, while 
observations are in IR. Dependence of $R$ on $z_t$ and $\lm$ will lead to a
drift of the star's image across the CCD chip during one exposure. 
This fact puts some limits on the exposure times and was investigated by
\citet[][]{roe02} for the Keck II telescope. If the exposure time
is too long, not only the measurement of position but also the
refraction correction is more uncertain.

\section{Dependence of AR on weather conditions and observing wavelength}

We derived a simplified method which will be further called a 
{\it semi-full} approach. In order to compute the monochromatic refraction 
angle, the equations \ref{rel_nroe} and \ref{rel_Rmon} were used (Roe's model). 
The partial water vapor pressure was computed in the following way. 
The values of maximum water vapor pressure for 
a given temperature ($p_{w,max}(T)$) are presented in Table \ref{tab1}. 
A 5-th order polynomial was fitted to this data, with $rms \simeq 0.051$, 
to derive a relation $p_w = H\,p_{w,max}(T)$ where $H$ is humidity. The 
following grid of parameters was used: 
$p\,[hPa]=613.25, 813.25, 1013.25=p_s$; $H\,[\%]=0, 50, 100$; 
$z_1\,[^\circ] = 0, 20, 40, 60$; $z_{21} \equiv z_2 - z_1\,['']=1, 5, 15$. 
For every point of this grid $R_{mon}$ was calculated for the temperatures 
from the range of 223.15 -- 293.15 $[K]$ every 1$K$. The range of weather
conditions was chosen in order to simulate real conditions in many observatories
starting from high-altitude ones where temperatures and air pressure values 
are low. 

For the mean refraction, a moderate simplification of Stone's (1996) method was used. 
$A(\lm)$ and $E(\lm)$ were computed using the equations 23--26 from 
\citet{sto96}. As an approximation of the spectral energy distribution 
$S(\lm)$, we used spectra of black body in temperature of 7000 $K$, which 
corresponds to a F0 star. The transmission of a telescope optics and quantum 
efficiency of a detector were assumed to be constant across a given band. 
Instead of an exact filter transmission curve, we used a model of an ideal 
filter characterized by the central wavelength $\lm_c$, total bandwidth 
$\Delta\lm$ and constant transmission, which for modern filters is true 
down to a level of a few \%. We applied the data for {\it Palomar High 
Angular Resolution Camera} (PHARO) for the K filter \citep{hay01}. 
As will be shown below, this semi-full approach is appropriate for relative 
astrometry.

\begin{figure*}
   \centering
   \includegraphics[width=\textwidth]{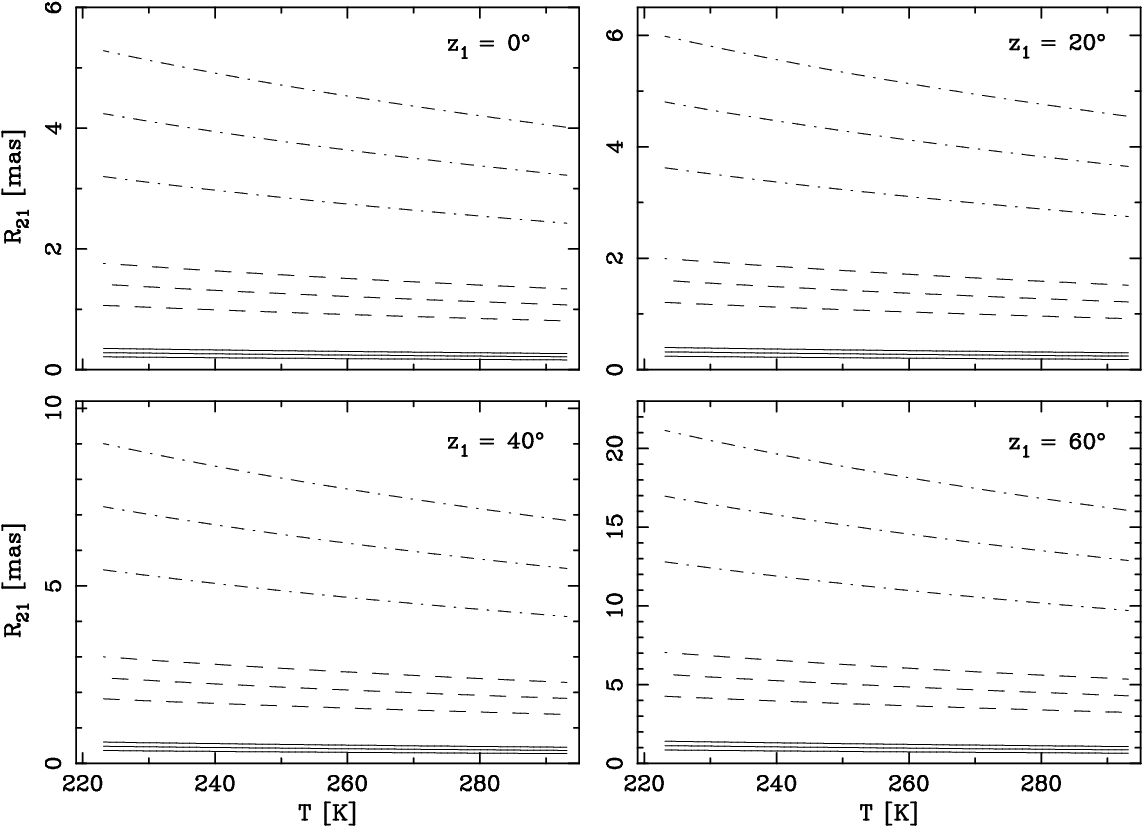}
      \caption{The refraction correction in the K band according to the
semi-full approach. Each  panel corresponds to a different zenithal distance of 0, 
20, 40 and 60 degrees. The solid lines are for $z_2 - z_1 =1\,$, dashed for 5 and 
dot-dashed for 15''. Every three lines correspond to $p=613.25\,hPa$ (bottom of 
the three), 813.25 $hPa$ (middle) and 1013.25 $hPa = p_s$ (top).\label{fig_Ktemp}}
   \end{figure*}

Figure \ref{fig_Ktemp} shows $R_{12}$ computed with the semi-full approach for 
the K band. The refraction correction is comparable to or bigger than 1 $mas$. 
Such a precision of relative astrometry is easily achievable from the ground. 
As the temperature and pressure dependence shows, the effect can also vary by 
a value higher than 1 $mas$. Obviously, the influence of the weather 
conditions decreases when the separation and zenithal distances are smaller. 
Less obvious is the fact that AR is more significant in low temperatures 
typical for high-altitude observatories. The correction $R_{21}$ is then 
about 30-40\% higher, the function is slightly steeper and the air pressure 
has an impact as well. Thus, one may say that it is better to 
observe close pairs, high over the horizon, in low air pressure and high 
temperatures. However high temperatures are an issue in the K band for two reasons. 
Firstly, more turbulence is created, so the AO correction is less efficient, 
thus the precision of the astrometry is lower. Secondly, the thermal background 
becomes more significant and variable. This means more difficulties in measuring 
accurate positions of stars in an image. 

\begin{figure*}
   \centering
   \includegraphics[width=\textwidth]{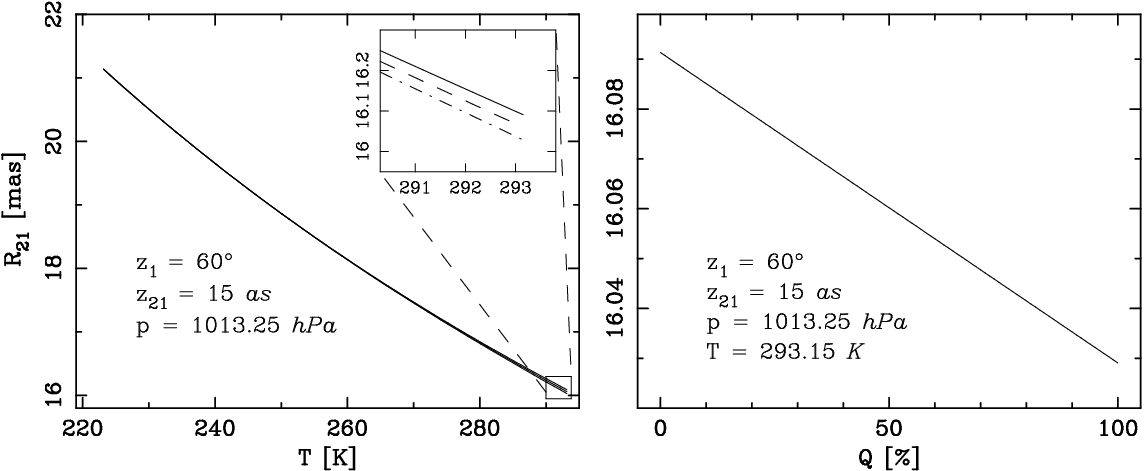}
      \caption{The impact of the humidity on the refraction correction (K band, 
semi-full approach). The solid line in the left panel refers to 0\% humidity, 
dashed to 50\% and dot-dashed to 100\% humidity. The right panel shows the 
dependence of $|R_{21}|$ on humidity at $T=293.15\,K$. \label{fig_Khum}}
   \end{figure*}

The partial water pressure (or humidity) is not as important in the semi-full
approach. Fig. \ref{fig_Khum} shows how the refraction correction changes with 
humidity. In the most extreme case (high $z$, $z_{21}$, air pressure and 
temperature), the scale of the change is smaller than 100 $\mu as$. This is a level 
of precision which can be achieved today \citep{hel08,neu06}. However, for more probable 
temperatures, lower pressure and smaller zenithal distance and separation, the 
typical refraction correction is much smaller. What is interesting, $R_{21}$ 
decreases as the humidity rises, and the steepness of the function is constant. 
If humidity would not be included into calculations (the last factor in 
Equation \ref{rel_nroe} would be 0), no significant 
uncertainties would occur but we recommend to keep this factor in mind when 
calculating AR corrections. 

\section{Requirements}

The results obtained by the computations of the refractive corrections 
with the semi-full approach were compared to calculations based on full 
computations of all terms in Eq. \ref{rel_sto} and with monochromatic 
refraction $R_{mon}$ itself. For the {\it full approach} the Mathar's model 
of the refractive index was used, $A(\lm)$ and $E(\lm)$ were computed as 
previously, a black-body spectrum with $T_{eff}\sim 7000\,K$ was used for both 
stars' $S(\lm)$. The transmission curve of PHARO camera optics, as well as 
PHARO's filters transmission curves (Fig. \ref{fig_n}) and the detector's quantum efficiency 
were adopted for $L(\lm)$, $F(\lm)$ and $D(\lm)$ respectively. This data 
was kindly sent by Dr. Bernhard Brandl from Leiden Observatory. The telescope 
optics transmission curve unfortunately was unavailable, thus it was assumed 
to be constant.

\begin{figure*}
   \centering
   \includegraphics[height=\textwidth, angle=270]{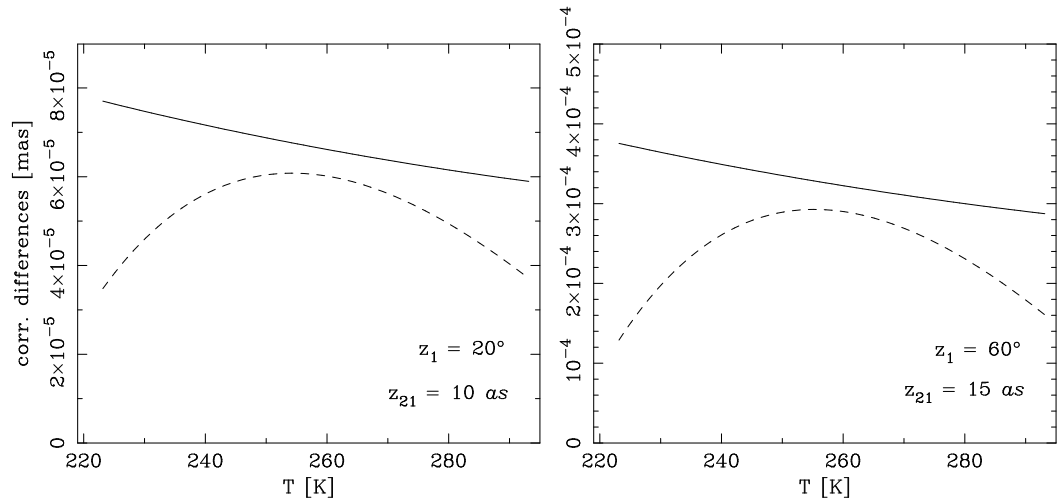}
	\caption{The differences between the monochromatic, $R_{21,mon}$, 
semi-full, $R_{21,sf}$ and full, $R_{21,f}$, computation of the refraction 
correction in the K band for F0 + F0 pair. $p=p_s$, $H=50\%$ and  
$z_1=20^{\circ}$, $z_{21}=10''$ (left) or $z_1=60^{\circ}$, $z_{21}=15''$
(right). The difference $R_{21,sf} - R_{21,mon}$ is denoted with the solid 
line and $R_{21,f} - R_{21,mon}$ withe the dashed line. \label{fig_sf}}
\end{figure*}

In Figure \ref{fig_sf} the differences between the monochromatic $R_{21,mon}$, 
semi-full $R_{21,sf}$ and full $R_{21,f}$ computations of the refraction corrections 
in the K band are shown. Humidity is set to 50\%, air pressure to $p_s$. $z_1-z_2$ is 
set to 10'' and $z_1$ to $20^{\circ}$ (left panel) or $z_1-z_2=15''$ 
and $z_1 = 60^{\circ}$ (right panel). They clearly show that the calculation of 
monochromatic AR only is good enough for achieving precision of single 
$\mu as$ in many probable sets of weather and observing conditions. For 
todays and future astrometric research in small fields in the infrared, one can 
compute only $R_{mon}$ set at $\lm_c$ of a certain filter. Knowledge of all
the transmission and quantum efficiency curves seems unnecessary, especially for 
the longer wavelengths.

Situation is rather more complicated when considering spectral types of stars. 
For stars with similar temperatures it does not matter if one uses full or 
semi-full approach. However when the effective temperatures differ, the 
differences between $R_{21,f}$ and $R_{21,sf}$ are not negligible if  $\mu as$ or 
even $mas$ precision is required. Due to a weak wavelength dependence, 
these differences are lower in K than in J band. 

\begin{figure*}
   \centering
   \includegraphics[width=\textwidth]{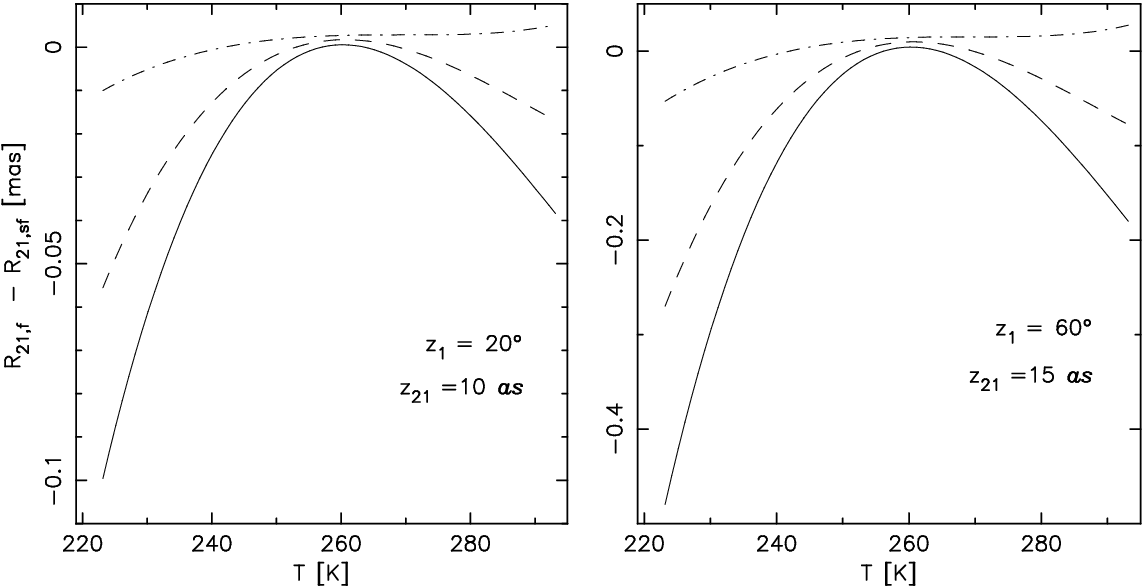}
	\caption{The differences between the full and semi-full 
($R_{21,f} - R_{21,sf}$), computation of the refraction correction in the 
K band for O8 + M1 (solid), M1 + M7(dashed), and F0 + F0 (dot-dashed line).
Observing conditions are given. $p = p_s$ and $H=50\%$. 
\label{fig_spT}}
\end{figure*}

The differences between the full and the semi-full approach in the K band are 
shown in Figure \ref{fig_spT} for the same conditions as in Fig. \ref{fig_sf}
for three systems -- O8 ($T_{eff}\simeq 37000\,K$) +  
M1 ($T_{eff}\simeq 3700\,K$), M1 + M7 ($T_{eff}\simeq 2700\,K$) and F0 + F0
($T_{eff}\simeq 7000\,K$). Brighter (hotter) stars are closer to the 
zenith. At the higher zenithal distance, for the first binary, errors caused by the 
usage of the semi-full approach do not allow to go down to a precision of 100 
$\mu as$ except for the narrow range of temperatures around 250 $K$. Situation
is only a bit better for the M-type binary. For the pair of identical stars,
this level of precision is easily achievable with the semi-full approach.

For a comparison, we also show the results of calculations for the same 
pairs of stars but observed higher over the horizon ($z_1 = 20^\circ$) and 
with smaller separation ($z_{21} = 10\, as$). For the new conditions, the 
situation in the K band improves (as expected) and even 10 $\mu as$ precision
is achievable in some conditions for every pair (especially around 250 $K$). It is 
also worthwhile mentioning that in the case of real astrometric measurements of 
O8 + M1 pair, an observer has to face several other sources of uncertainties 
caused by the large contrast of the observed stars. 
In the J band, the results would reveal a more complicated behavior.
For cooler stars (or rather black-bodies) the maximum of the 
spectral energy distribution lies in the neighborhood of the J band. 
At 1.25 $\mu m$ different types of stars have different shape of $S(\lm)$, 
while at 2.2 $\mu m$ the energy distribution looks more or less similar, no 
matter what the spectral type is. One must also remember that the 
refraction index $n$ is a steeper function of $\lm$ in J than in K band. 
Thus it is better to calculate AR in the K than in J band or simply observe 
with a narrow-band filter. 

This may be still not enough in the case of exoplanets. Their spectra are
extremely different from the black-body. Possible various abundances of 
molecules like water, methane, ozone, carbon dioxide and many others would 
make every planet's spectrum unique. After direct imaging of a planet, 
which is said to be less difficult with ELTs, precise astrometry may be 
possible only after obtaining a high signal to noise spectrum in IR.

\begin{figure*}
   \includegraphics[width=\textwidth]{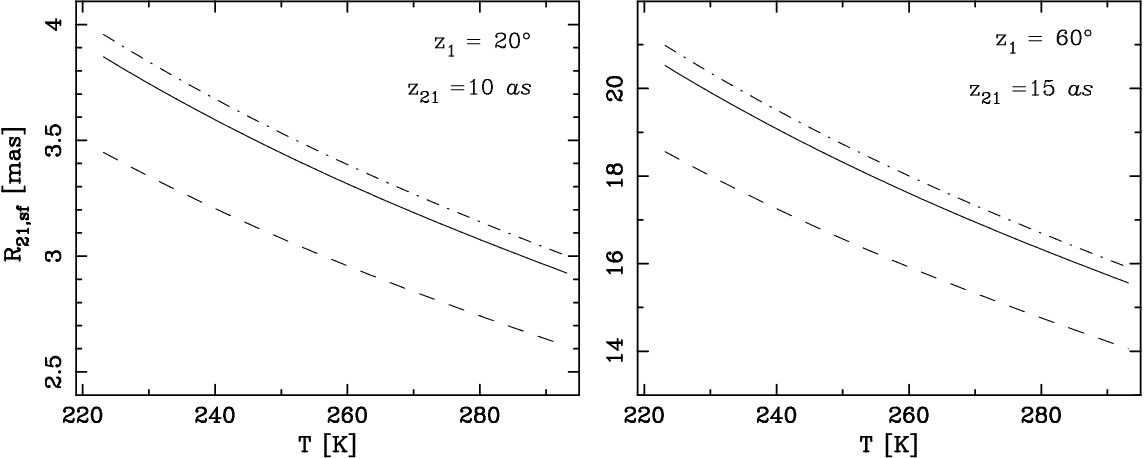}
	\caption{The dependence of the refraction correction in semi-full 
approach $R_{21,sf}$ on the filter bandwidth for K band.
Calculations are for O8 + M1 pair. Observing conditions are given, $p=p_s$ 
and $H=50\%$. In both cases solid lines denote ''normal'' filter ($\lm_c$, 
$\Delta\lm$), dashed line -- ''wide'' filter ($\lm_c$, $2 \Delta\lm$) and
dot-dashed -- ''narrow'' filter ($\lm_c$, $\frac{1}{2} \Delta\lm$). 
\label{fig_band}}
\end{figure*}

The influence of bandwidth on the refraction correction in the semi-full 
nomenclature is shown in Figure \ref{fig_band}. Once again we used O8 + M1 pair 
observed $20^\circ$ from zenith, separated by 10 $mas$ (left) and observed 
$60^\circ$ from zenith, separated by 15 $mas$ (right). The air pressure was set 
to $p_s$ and 50 \% humidity was assumed. The plot shows the refraction correction 
$R_{21,sf}$for: 1) a perfect filter with the central wavelength $\lm_c=2.196\, 
\mu m$ and bandwidth $\Delta\lm=0.336\, \mu m$ that corresponds to PHARO's 
K filter (solid line) -- ''normal'', 2) a ''wide'' filter with the same 
central wavelength but 2 times wider bandwidth (dashed line), 3) and a 
''narrow'' filter with the same $\lm_c$ but 2 times smaller bandwidth. 
When $\Delta\lm$ for a given band reaches zero, $R_{21,sf}$ becomes $R_{21,mon}$.
The correction is the smallest in ''wide'' filter and the biggest in ''narrow'' one,
but only if the hotter star is higher over the horizon. It is exactly the opposite
when M1 star is closer to the zenith. The differences between these two cases are 
the smallest in the "narrow".case.

As seen in Figures \ref{fig_Ktemp}, \ref{fig_sf} and \ref{fig_spT}, the required 
precision of air pressure and temperature readings may vary with the zenithal distance 
and separation of stars. Especially the dependence on temperature in the full 
approach is interesting due to the influence of the resonances near the K band. 
For close pairs high over the horizon, any reasonable temperature and pressure 
can be set and $mas$ precision is reachable. Note that in the semi-full 
approach even at 60 degrees from zenith, $p=p_s$, $R_{21}$ changes from 
1.4125 to 1.0732 $mas$ across a given temperature range which gives about 4.85 
$\mu as$ per 1 $K$. For the same case, at the constant temperature 223.15 $K$, 
the refraction correction changes from 0.8551 at 613.25 $hPa$ to 1.4125 $mas$ at 
$p_s$ which gives about 1.39 $\mu as$ per 1 $hPa$. This allows us to achieve 
1 $mas$ actually without any knowledge of weather readings and 10 $\mu as$ 
precision  when conditions are known with an uncertainty of 1 unit. At the same time, 
it suggests that for other cases, when $R_{21}$ is bigger, a much higher than 1 
unit precision of weather condition readings might be required to achieve 1 $\mu as$.

In Table \ref{tab2} all the requirements for high-precision astrometry in
the K band are collected. The maximum allowable errors of weather conditions 
readings are given and the need of knowing spectral type of stars (Sp.T.) 
and usage of the full approach (F.Ap.) is specified\footnote{Note, that when 
sp. type is needed, the usage of monochromatic refraction only is forbidden}. 
The precision in humidity is not specified as a typical 1\% precision will 
affect astrometry at a level comparable to 1 $\mu as$ in the most extreme cases.
For a given zenithal distance and $z_{21}$, the set of requirements 
will lead to measurement errors at the level of given precision $\sigma$ 
(in Table \ref{tab2}) or significantly smaller. As one can see, to reach the 
$\mu as$ level precise readings of air temperature and pressure are 
needed in some cases. Those numbers can be compared to the accuracies of 
real measurements. For example the weather station at Paranal, provides the 
following accuracies of readings: 0.2 $K$ in temperature, 0.1 $hPa$ in pressure 
and 1\% in humidity. It simply means that 10 $\mu as$ is not achievable 
there.

Unfortunately, milikelvin or higher variations of temperature as well as variations 
of the air pressure around the telescope's dome are very probable. This means that 
achieving $\mu as$-measurements in wider fields can be impossible, at least by using 
this model of Earth's atmosphere, even if weather instruments would be accurate enough. 
It refers not only to single-mirror telescopes but also to interferometers. Ground based 
astrometry might be limited to 1 or even 10 microarcsecond level by the impossibility  
to carry out proper AR calculations.

\section{Summary}
Undoubtedly, it is necessary to account for the atmospheric refraction to achieve 
precision of astrometric measurements at the level of miliarcseconds or 
better. In order to simplify the procedure it is better to observe in the infrared 
(the longer wavelength, the better) and with narrow-band filters. In cases of 
the most reasonable zenithal distances and observing conditions, the usage of the 
semi-full approach (without the transmission and quantum efficiency curves and 
with the black-body spectrum, Roe's refraction model) allows to obtain precision 
well below 1 miliarcsecond. In many cases, also the monochromatic 
refraction is sufficient. Nevertheless, to  compute AR properly one still has to 
know the air temperature and pressure rather well. It is also good to estimate 
stars' spectral types when $R_{mon}$ is not enough. 
In the semi-full approach (or any other) standard precision of weather readings 
(0.1 $K$, 0.01 $hPa$ and 1\% in humidity) is enough to reach 
$\sigma \sim 10 \mu as$ which is still mostly unreachable by todays facilities. 
Nevertheless, future ELT-s will probably be more accurate astrometrically but 
this will require more accurate weather readings and maybe a better model of
the Earth's atmosphere. Direct measurements of $n$ in various conditions are
usually carried out in laboratories or over long distances but inside the troposphere, 
so the light beam travels through a relatively stable environment, which is not the case 
in astronomy. Thus an improved atmosphere model, which may be required for reaching 
a 1 - 10 $\mu as$ level of precision, would have to be calibrated using the
measurements of stars' positions taken from the space, which will be achievable only 
after launching {\it Gaia} or {\it SIM}.

\section*{Acknowledgements}
K.G.H. would like to thank Prof. Maciej Konacki from the Nicolaus Copernicus 
Astronomical Center in Toru\n\ for discussions and useful advices, 
Dr. Bernhard Brandl from the Leiden Observatory for sending the transmission 
and efficiency data of PHARO camera, Dr. Thomas Bertram from the Faculty of 
Mathematics and Natural Sciences of University of Cologne for valuable corrections,
Prof. Andrzej A. Marsz from the Gdynia Maritime University for making 
available the data for Table \ref{tab1}, Dr. Marc Sarazin from ESO for the 
information on the VLT observatory weather station, and an anonymous referee
for usefull advices and priceless suggestions.

This work was supported by the 
Polish Ministry of Science and Higher Education through grants 
N203 005 32/0449 and 1P03D-021-29 and by the Foundation for Polish Science
through a FOCUS grant.

\clearpage
\clearpage
\clearpage
\clearpage
\clearpage

\clearpage

\clearpage

\clearpage
\clearpage

\begin{table}
\caption{Maximum water vapor pressure as a function of the temperature.\label{tab1}}
\centering
\begin{tabular}{cc|cc}
\hline\hline
Temper. & $p_{w,max}$ & Temper. & $p_{w,max}$\\
$[^\circ C]$ & $[hPa]$ & $[^\circ C]$ & $[hPa]$\\
\hline
50&      123.3 & 0 &     6.11\\
45&      95.77 &-5 &     4.21\\
40&      73.72 &-10&     2.68\\
35&      56.2  &-15&     1.9\\
30&      42.41 &-20&     1.25\\
25&      31.66 &-25&     0.8\\
20&      23.27 &-30&     0.5\\
15&      17.05 &-35&     0.309\\
10&      12.28 &-40&     0.185\\
5 &      8.72  &-45&     0.108\\
\hline
\end{tabular}
\end{table}

\begin{table}
\caption{The weather reading and other requirements for a relative astrometry 
with a given precision. The uncertainty in the humidity is assumed to be 1\%. 
The symbols $dT$ and $dp$ refer to the errors in the temperature and pressure 
corresponding to an error in the refraction correction at least two times 
smaller than a given precision $\sigma$. The symbol ''$n-n$'' stands for 
{\it not necessary} --- meaning that across a reasonable range of a parameter 
the variations in $R_{21}$ are at least 2 times smaller than a given 
astrometric precision.\label{tab2}}
\centering
\scriptsize
\begin{tabular}{c|cccc|cccc}
\hline\hline
$dz$ & $dT$ & $dp$ & Sp.T & F.Ap. & $dT$ & $dp$ & Sp.T. & F.Ap. \\
$[as]$ & $[K]$ & $[hPa]$ &  &  & $[K]$ & $[hPa]$ &  &  \\
\hline
\multicolumn{9}{l}{$\sigma \sim 1\, mas$:}\\
\hline
& \multicolumn{3}{l}{$z=0^{\circ}$} && \multicolumn{4}{l}{$z=20^{\circ}$}\\
1  & n-n & n-n & no & no & n-n & n-n & no & no \\
5  & n-n & 100 & no & no & n-n & 100 & no & no \\
15 & 10 & 50 & no & no & 1 & 50 & no & no \\
\hline
& \multicolumn{3}{l}{$z=40^{\circ}$} && \multicolumn{4}{l}{$z=60^{\circ}$}\\
1  & n-n & n-n & no& no& n-n & 100 & yes & no \\ 
5  & 10 & 100 & no & no & 10 & 50 & yes & no \\
15 & 10 & 50 & no & no & 5 & 10 & yes & no \\
\hline
\multicolumn{9}{l}{$\sigma \sim 100 \mu as$:}\\
\hline
& \multicolumn{3}{l}{$z=0^{\circ}$} && \multicolumn{4}{l}{$z=20^{\circ}$}\\
1  & 10 & 100 & no & no & 10 & 100 & no & no \\
5  & 10 & 10 & no & no & 5 & 10 & no & no \\
15 & 1 & 5 & no & no & 0.5 & 5 & yes & no \\
\hline
& \multicolumn{3}{l}{$z=40^{\circ}$} && \multicolumn{4}{l}{$z=60^{\circ}$}\\
1  & 10 & 50 & no & no & 10 & 10 & yes & no \\ 
5  & 1 & 10 & yes & no & 5 & 10 & yes & no \\
15 & 1 & 5 & yes & no & 1 & 5 & yes & yes \\
\hline
\multicolumn{9}{l}{$\sigma \sim 10\, \mu as$:}\\
\hline
& \multicolumn{3}{l}{$z=0^{\circ}$} && \multicolumn{4}{l}{$z=20^{\circ}$}\\
1  & 1 & 10 & no & no & 1 & 10 & yes & no \\
5  & 1 & 1 & no & no & 0.5 & 1 & yes & no \\
15 & 0.1 & 1 & no & no & 0.1 & 0.5 & yes & no \\
\hline
& \multicolumn{3}{l}{$z=40^{\circ}$} && \multicolumn{4}{l}{$z=60^{\circ}$}\\
1  & 1 & 5 & yes & no & 1 & 1 & yes & no \\
5  & 0.1 & 1 & yes & no & 0.1 & 0.5 & yes & no \\
15 & 0.1 & 0.5 & yes & no & 0.05 & 0.1 & yes & yes \\
\hline
\multicolumn{9}{l}{$\sigma \sim 1\, \mu as$:}\\
\hline
& \multicolumn{3}{l}{$z=0^{\circ}$} && \multicolumn{4}{l}{$z=20^{\circ}$}\\
1  & 0.1 & 1 & no & no & 0.1 & 1 & yes & no \\
5  & 0.1 & 0.1 & no & no & 0.1 & 0.1 & yes & yes \\
15 & 0.01 & 0.1 & no & yes & 0.01 & 0.05 & yes & yes \\
\hline
& \multicolumn{3}{l}{$z=40^{\circ}$} && \multicolumn{4}{l}{$z=60^{\circ}$}\\
1  & 0.1 & 0.5 & yes & no & 0.1 & 0.1 & yes & no \\
5  & 0.01 & 0.1 & yes & yes & 0.01 & 0.05 & yes & yes \\
15 & 0.01 & 0.05 & yes & yes & 0.005 & 0.01 & yes & yes \\
\hline
\end{tabular}
\end{table}

\end{document}